# Geminate recombination of hydroxyl radicals generated in 200 nm photodissociation of aqueous hydrogen peroxide. [1]

Robert A. Crowell,  Rui Lian, M. C. Sauer, Jr., Dmitri A. Oulianov, and Ilya A. Shkrob [*]

*Chemistry Division , Argonne National Laboratory, Argonne, IL 60439*

**Abstract**

The picosecond dynamics of hydroxyl radicals generated in 200 nm photoinduced dissociation of aqueous hydrogen peroxide have been observed through their transient absorbance at 266 nm. It is shown that these kinetics are nearly exponential, with a decay time of ca. 30 ps. The prompt quantum yield for the decomposition of $H_2O_2$ is 0.56, and the fraction of hydroxyl radicals escaping from the solvent cage to the water bulk is 64-68%. These recombination kinetics suggest strong caging of the geminate hydroxyl radicals by water. Phenomenologically, these kinetics may be rationalized in terms of the diffusion of hydroxide radicals out of a shallow potential well (a solvent cage) with an Onsager radius of 0.24 nm.

___________________________________________________________

## 1. Introduction

UV-light induced photodissociation of aqueous hydrogen peroxide

$$H_2O_2 \xrightarrow{h\nu} 2\ HO \qquad (1)$$

is widely used in water treatment, medical equipment sterilization, bleaching, etc. [1,2] These practical uses are facilitated by the high oxidation potential of the photogenerated hydroxyl radicals and large primary quantum yield of $H_2O_2$ decomposition, 0.4-0.5 [3,4]. The overall quantum yield is further increased to 1-2, due to Haber-Weiss chain reactions [1,5]

$$HO + H_2O_2 \longrightarrow H_2O + HO_2 \qquad (2)$$

$$HO_2 + H_2O_2 \longrightarrow O_2 + H_2O + HO \qquad (3)$$

with rate constants of $2.7 \times 10^7$ $M^{-1}$ $s^{-1}$ and $7 \times 10^9$ $M^{-1}$ $s^{-1}$, respectively [5]. Reaction (2) competes with the diffusion-controlled recombination

$$HO + HO \longrightarrow H_2O_2 \qquad (4)$$

that proceeds with a rate constant $k_4 = 4.7 \times 10^9$ $M^{-1}$ $s^{-1}$ at 25°C [5].

While the general photochemistry of $H_2O_2$ decomposition is well understood, no mechanistic insight yet exists as to the dynamics of geminate (HO..OH) pairs generated in photoreaction (1). In this Letter, the dynamics of these hydroxyl radicals using 200 nm pump - 266 nm probe ultrafast transient absorption spectroscopy are reported. We also report an improved measurement for the primary quantum yield of free hydroxyl radicals in 248 nm photodissociation of $H_2O_2$.

## 2. Experimental

*Picosecond pump-probe kinetics.* The ultrafast kinetic measurements reported below were carried out using a 1 kHz Ti:sapphire setup. A diode-pumped Nd:YVO laser was used to pump Kerr lens mode-locked Ti:sapphire laser operating at 80 MHz (Spectra

Physics Tsunami). The 45 fs FWHM pulses from the oscillator where stretched to 80 ps in a single grating stretcher, and single pulses were then selected at 1 KHz with a Pockels cell. The 2 nJ pulses were amplified to 4 mJ in a two-stage multipass Ti:sapphire amplifier and passed through a grating compressor that yielded Gaussian probe pulses of 60 fs FWHM and 3 mJ centered at 800 nm. The amplified beam was split into three parts. The first 800 nm beam was used to generate 200 nm (fourth harmonic) pulses used as a pump, as explained below: it was passed through two tandem BBO crystals that generated the second (Type I, 500 μm, 29°) and third (Type II, 100 μm, 58°) harmonic. The fourth harmonic is produced by upconversion of the third harmonic (266 nm) and the second 800 nm beam in a third BBO crystal (Type I, 100 μm, 68°). Up to 20 μJ of the 200 nm light was produced this way (300 fs FWHM pulse). The pump power, before and after the sample, was determined using a calibrated power meter (Ophir Optronics model 2A-SH). The 266 nm probe was produced by third harmonic generation with two 200 μm thick BBO crystals using the third 800 nm beam. To remove the 1st and 2nd harmonics, the resulting beam was passed through a 3 nm FWHM notch filter and then reflected, in succession, from four 266 nm dielectric mirrors placed in each arm of the spectrometer. Test experiments showed that no residual signal from the lower harmonics reached the photodiode detectors.

The pump and probe beams were perpendicularly polarized, focused to round spots of 100 μm and 20 μm radius, respectively, and overlapped in the sample at 6.5°. The spot sizes were determined by scanning the beam with a 10 μm pinhole. The signal and the reference probe pulses (< 1 nJ) were detected with fast silicon photodiodes, amplified, and sampled using home-built sample-and-hold electronics and a 16 bit A/D converter. A mechanical chopper locked at half the repetition rate of the laser was used to block the pump pulses on alternative shots. No dependence of the transient absorbance on the pump and probe polarization was found.

Aqueous $H_2O_2$ solution was prepared by mixing 35 wt% hydrogen peroxide (Aldrich; stabilized with a trace of tin) with deaerated nanopure water. The solution was constantly purged with dry nitrogen. The experiments were carried out using a 150 μm thick, 6 mm wide high-speed (10 m/s) jet with a stainless steel nozzle. An all 316 stainless steel and Teflon flow system was used to pump the solution through this jet at 0.5-0.75 dm$^3$/min.

Importantly, two photon photoionization of the solvent (water) is negligible under the conditions of our experiment because the two-photon absorbance of 200 nm light by water is orders of magnitude lower than one-photon absorbance by 1 M $H_2O_2$ (OD$_{200}$ across the jet is ca. 3). Even without 1 M $H_2O_2$ in the solution, the 266 nm transient absorption signal from the hydroxyl radical and hydrated electron generated in the course of the photoionization was < 1 % of the signal from the hydrogen peroxide solution, under identical irradiation conditions.

*Nanosecond kinetics.* Fifteen nanosecond fwhm, 1-20 mJ pulse from an KrF excimer laser (Lamda Physik LPX 120i) was used to photolyze $N_2$-saturated aqueous solutions containing 20-40 mM of $H_2O_2$ and 1 M $NaHCO_3$. The absorbance of the 248 nm light by the bicarbonate was negligible (< 0.021 M$^{-1}$ cm$^{-1}$). 0.2 L of the solution was circulated through the cell using a peristaltic pump. The 1.36 mm optical path cell had 1 mm thick suprasil windows. The laser beam was masked with a 3 mm x 6 mm rectangular brass aperture. This beam was normal to the windows; the analyzing light (> 500 nm) from a superpulsed Xe lamp was crossed at 30$^o$ with this 248 nm excitation beam. The wavelength of the analyzing light (600 nm) was selected using a 10 nm fwhm narrow band interference filter.

A fast silicon photodiode (EG&G model FND100Q, biased at -90 V) with a 1.2 GHz video amplifier (Comlinear model CLC100) terminated into a digital signal analyzer (Tektronix model DSA601) were used to sample the transient absorbance kinetics (3 ns response time). Two calibrated energy meters (Molectron model J25-080) were used to measure the power of the incident and transmitted 248 nm light.

**3. Results and Discussion.**

*200 nm pump-probe kinetics.* Fig. 1 shows transient absorption kinetics of hydroxyl radicals observed at 266 nm in 200 nm photolysis of 1 M $H_2O_2$. The short-lived "spike" observed within the duration of 200 nm light pulse is due to simultaneous absorption of both the pump and probe pulses (i. e., "1+1" nonlinear absorbance) and gives the instrument response function; this "spike" is a common artifact in UV pump-probe studies [6].

As seen from Fig. 1, the kinetics observed after the excitation pulse are nearly exponential. The hydroxyl radicals rapidly escape from the solvent cage into the water bulk (with a time constant of 29 ps); from the exponential fit, the escape fraction of these radicals is ca. 68%. To estimate the quantum yield of photoreaction (1), the following approach was used: The radial profiles for the pump and probe beams were assumed to be Gaussian, so that incident (time-integrated) photon fluence for the pump beam was given by

$$J_{pump}(r) = J_0 \exp(-[r/r_{pump}]^2) \tag{5}$$

and the beam power was $I_{pump} = \pi r_{pump}^2 J_0$. As the pump beam penetrates the sample, it is attenuated as $exp(-\beta x)$, where $\beta$ is the absorption coefficient of the sample and $x$ is the penetration depth. Neglecting the absorbance of 200 nm light by photogenerated hydroxyl radicals, the total number of 200 nm photons absorbed by the sample of thickness $L$ is given by $I_{abs} = I_{pump}(1 - \exp(-\beta L))$. The sample-average yield $P(r)$ of a photoproduct across this sample is given by equation

$$P(r) = \phi\, J_{pump}(r)\,(1 - \exp(-\beta L))/L \tag{6}$$

where $\phi$ is the quantum yield for the photoproduct. Assuming that the absorption of the probe light is weak, $P(r)L\varepsilon_{probe} \ll 1$ (which was the case in our experiment), where $\varepsilon_{probe}$ is the extinction coefficient of the photoproduct (hydroxyl radical) at the probe wavelength of 266 nm, the loss $\Delta J_{probe}(r)$ in the intensity of the 266 nm probe light is given by equation

$$\Delta J_{probe}(r) = -J_{probe}(r) P(r) \varepsilon_{probe} L \tag{7}$$

where $J_{probe}(r)$ is the radial distribution function of the probe light given by eq. (5) with the beam radius of $r_{probe}$. Combining all of these equations, and carrying out the integration over $r$, the beam-average loss in the probe transmission defined as

$$-\Delta T/T = \int_0^\infty dr\ 2\pi r\ \Delta J_{probe}(r) \bigg/ \int_0^\infty dr\ 2\pi r\ J_{probe}(r) \qquad (8)$$

is given by

$$-\Delta T/T = -\varepsilon_{probe}\ \phi\ (1-\exp(-\beta L))\ I_{pump} \big/ \left[\pi\left(r_{pump}^2 + r_{probe}^2\right)\right] \qquad (9)$$

From eq. (9) we obtain a general formula

$$\phi = \left([-\Delta T/T]/\varepsilon_{probe}\right) \bigg/ \left(I_{abs} \bigg/ \pi\left[\left(r_{pump}^2 + r_{probe}^2\right)\right]\right) \qquad (10)$$

Thus, to obtain $\phi$ using eq. (10), the sample thickness $L$ and the absorption coefficient $\beta$ of the sample at the excitation wavelength need not be known; it is sufficient to determine the optical density for the probe light and the pump power before and after the sample. Note that the prompt quantum yield for the decomposition of $H_2O_2$ is 1/2 of the quantum yield of hydroxyl radicals.

For the kinetic trace shown in Fig. 1, $I_{pump}$=3 µJ, $r_{pump}$= 100 µm and $r_{probe}$=20 µm. The molar extinction coefficients $\varepsilon_{probe}$ for hydroxyl radicals at 266 nm are not known accurately: the estimates vary by as much as 25%, from 370 to 465 M$^{-1}$ cm$^{-1}$ [7-12]. Since these coefficients were determined in pulse radiolysis experiments, these coefficients depend on the estimates for the radiolytic yields; the latter have been revised repeatedly after the original experiments were carried out [5-13]. Perhaps, the most reliable measurement is by Nielsen et al. [8], $\varepsilon_{probe}$ =420 M$^{-1}$ cm$^{-1}$. Using this estimate, the average concentration of the hydroxyl radicals across the jet was 0.8 mM and the quantum yield of peroxide decomposition at 0 and 500 ps was 0.56 and 0.38, respectively. At 200 nm, hydrogen peroxide has an extinction coefficient of 200 M$^{-1}$ cm$^{-1}$ [8] and the pump light was absorbed in 22 µm (i.e., all of the 200 nm light was absorbed by the jet), so that the concentration of radicals near the jet surface was 5.6 mM. While this concentration is relatively high, given the rate of reaction (4) the lifetime of free hydroxyl radical was 40 ns, i.e., no cross recombination occurred within the observation window (500 ps). Reaction (2) is also too slow ($k_2$=2.9x10$^7$ M$^{-1}$ s$^{-1}$ [5]) to affect the geminate dynamics of hydroxyl radicals in 1 M hydrogen peroxide.

Our estimate of 0.38 for the primary quantum yield for decomposition of $H_2O_2$ (after the escape of hydroxide radicals to the bulk) compares favorably with the estimates given in the literature: 0.49±0.07 [3] and 0.47±0.03 [4] at 254 nm and 0.45±0.06 at 222 nm [2] (also, see below). This estimate can be brought into a better agreement with these previous measurements by reconsideration of the extinction coefficient for the hydroxyl radical. The radiolytic measurements of Nielsen et al. [8] and Czapski and Bielski [9] relied on the yield of 5.3 hydroxyl radicals per 100 eV of radiation absorbed by $N_2O$-saturated water, whereas other work suggests a 15% higher yield (6.1 per 100 eV) [13]. This readjustment decreases the molar absorptivity to 400 M$^{-1}$ cm$^{-1}$ (which also compares better to 370 M$^{-1}$ cm$^{-1}$ obtained in the photolytic experiments of Boyle et al [10]) and increases our estimate of the quantum yield from 0.38 to 0.44 - in agreement with the previous work [2,3,4] and the estimates given below (in such a case, the prompt quantum yield of 0.56 should be increased to 0.65).

*Quantum yield for 248 nm photolysis.* All of the previous estimates for the quantum yield of $H_2O_2$ decomposition were obtained by product analysis (evolution of gaseous oxygen) at the end of chain reactions (1) to (4). To improve on these indirect estimates for the yield of reaction (1), we used 248 nm laser flash photolysis to observe the primary yield of *free* hydroxyl radicals through their fast reaction with bicarbonate anions ($k_{11}$= 1.5x10$^7$ M$^{-1}$ s$^{-1}$ [14] or 8.5x10$^6$ M$^{-1}$ s$^{-1}$ [15] )

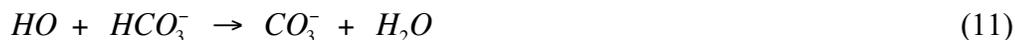

$$HO\ +\ HCO_3^-\ \rightarrow\ CO_3^-\ +\ H_2O \qquad (11)$$

At the *pH*=8.3 (1 M bicarbonate), all hydrogen peroxide ($pK_a$=11.7 [5]) is in the protonated form. The formation of the carbonate radical anion was observed at 600 nm ($\varepsilon_{600}(CO_3^-)$=1860 M$^{-1}$ cm$^{-1}$ [14]); this formation was complete in 200 ns (the apparent rate constant for reaction (11) was 2x10$^7$ M$^{-1}$ s$^{-1}$). The second-order decay of the carbonate radical anion ($2k$=1.25x10$^7$ M$^{-1}$ s$^{-1}$ [16]) was slow, with $t_{1/2}$ of 5-10 µs, so the loss of the 600 nm signal due to the cross recombination was negligible. The reaction of $CO_3^-$ with $H_2O_2$ is also very slow ($k=8 \times 10^5$ $M^{-1}$ $s^{-1}$ [16]). The highest yield of the hydroxyl radicals estimated from our kinetic data was 140 μM; at this concentration, all of the reactions competing with reaction (11) (such as rxn. (2) and (4)) are too slow to change the observed yield of $CO_3^-$ radical anions at $t=200$ ns. Note that 2 μM hydroxide and 9.9 mM carbonate are present in the solution due to the protic equilibria of the carbonate system. The hydroxide anion rapidly reacts with the hydroxyl radical yielding $O^-$ [5], but the concentration of the hydroxide is too low to affect our measurement. While the carbonate anion also reacts with the hydroxyl radical with rate constant of $4.2 \times 10^8$ $M^{-1}$ $s^{-1}$ [14], the product of this reaction is $CO_3^-$, i.e., no loss of the 600 nm absorbance from the carbonate radical anion ensues.

The transient absorbance at 600 nm (at $t=200$ ns) was plotted *vs.* the number of the absorbed photons. The initial slope of this curve (for 248 nm laser fluence < $10^{16}$ photons/cm$^2$) corresponds to a quantum yield of 0.44±0.01, in reasonable agreement with the previous measurements. At higher 248 nm laser fluence ( (2-to-6)x$10^{16}$ photons/cm$^2$), the $\Delta OD_{600}$ plot exhibited a negative curvature (for example, the apparent quantum yield decreases to 0.384 for the fluence of 4x$10^{16}$ photons/cm$^2$). This negative curvature is accounted for by the absorption of the 248 nm light by hydroxyl radicals that are generated within the duration of the laser pulse ($\varepsilon_{248}(HO)=500$ $M^{-1}$ $cm^{-1}$ [9]). The molar absorptivity of $H_2O_2$ at 248 nm is very low compared to that of the hydroxyl radicals formed in reaction (1) (we obtained $\varepsilon_{248}(H_2O_2)=26$ $M^{-1}$ $cm^{-1}$ using an Olis/Cary-14 spectrophotometer and 24.8 $M^{-1}$ $cm^{-1}$ from the KrF laser transmission data), i.e., under the conditions of our experiment up to 20-40% of the 248 nm photons are absorbed by these radicals. Numerical simulations indicate that this effect fully accounts both for the negative curvature of the $\Delta OD_{600}$ plot and the observed linear decrease in the transmission of the 248 nm light with the laser power. Note that this effect is minor in our picosecond experiment because the concentration of hydrogen peroxide ( 1 M vs. 20-40 mM) and its extinction coefficient at 200 nm (200 $M^{-1}$ $cm^{-1}$ vs. 26 $M^{-1}$ $cm^{-1}$) are both considerably higher. Thus, < 0.5 % of the 200 nm light was absorbed by the hydroxyl radicals, which justifies the use of eq. (10).

*Kinetic analysis.* The most intriguing of our observations is that the recombination kinetics of hydroxyl radicals are nearly exponential, whereas a very different dependence was expected for the geminate recombination controlled by free diffusion of these radicals. This suggests a weak interaction between the radicals (in other words, the existence of a solvent cage around the radical partners). Phenomenologically, this interaction can be described in terms of a mean force potential $U(r)$ having a profile of a well [17,18]. As shown by Shushin [17], the decay kinetics of diffusional escape from such a potential well exhibits two regimes: (i) an exponential decay on a short time scale (with rate constant $W=W_r+W_d$, where $W_{r,d}$ are the recombination and dissociation rates of the radicals in the potential well, respectively), and $t^{-1/2}$ behavior on the longer time scale, due to slow recombination of radicals that escaped beyond the Onsager radius $a$, at which $U(a)=-kT$. In this theory, the escape probability $p_d$ of the radical partners is equal to $W_r/W$ and the effective radius $R_{eff}$ of the reaction in the bulk is given by a product $a(1-p_d)$ [18]. Fig. 1 demonstrates the fit of the experimental kinetics to Shushin's theory expressions [17,18] for the survival probability of a radical pair migrating out of the potential well. Using a diffusion coefficient of $2.8\times10^{-5}$ cm$^2$/s for the hydroxyl radical [5], one obtains the parameters $p_d=0.64$, $a=0.24$ nm, and $W^{-1}=20.6$ ps, which in turn yield $R_{eff}=0.09$ nm. A direct estimate from the known rate constant of reaction (4) in the bulk [5] gives the recombination radius of 0.11 nm.

### 4. Conclusion.

Photoexcitation of hydrogen peroxide at 200 nm causes its rapid (< 300 fs) dissociation with the formation of weakly interacting geminate hydroxyl radicals in the solvent cage. At 25°C, the decay of these radicals from the cage takes ca. 20 ps, with 64% of the radicals escaping to the water bulk and the rest recombining. The prompt quantum yield for the photodissociation of $H_2O_2$ is 0.56 (or 0.65, depending on the assumed estimate for the molar absorptivity of hydroxyl radical; see above), and the Onsager radius of the attractive potential is 0.24 nm. The absolute primary quantum yield for $H_2O_2$ decomposition (to *free* hydroxyl radicals) for 248 nm photoexcitation is 0.44.

### 5. Acknowledgement.

This work was performed under the auspices of the Office of Science, Division of Chemical Science, US-DOE under contract number W-31-109-ENG-38. We thank Prof. S. E. Bradforth of USC and Dr. D. M. Bartels of NRDL for providing the motivation for this study. IAS thanks Dr. S. V. Lymar of BNL for the introduction to the perils of aqueous radiation chemistry.

Fig. 1.

Transient absorbance at 266 nm detected in 200 nm pulse excitation (300 fs FWHM, $I_{pump}$=3 µJ; $r_{pump}$=100 µm) of 1 M aqueous $H_2O_2$ in a 150 µm jet. The pump fluence was $9.5 \times 10^{-3}$ J/cm$^2$. This signal is from hydroxyl radicals, 64% of which escape the recombination in the cage. These decay kinetics were simulated using Shushin's semianalytical theory of diffusion in a shallow potential well (the solid line is the least-squares fit of the kinetic trace for $t$>3 ps).

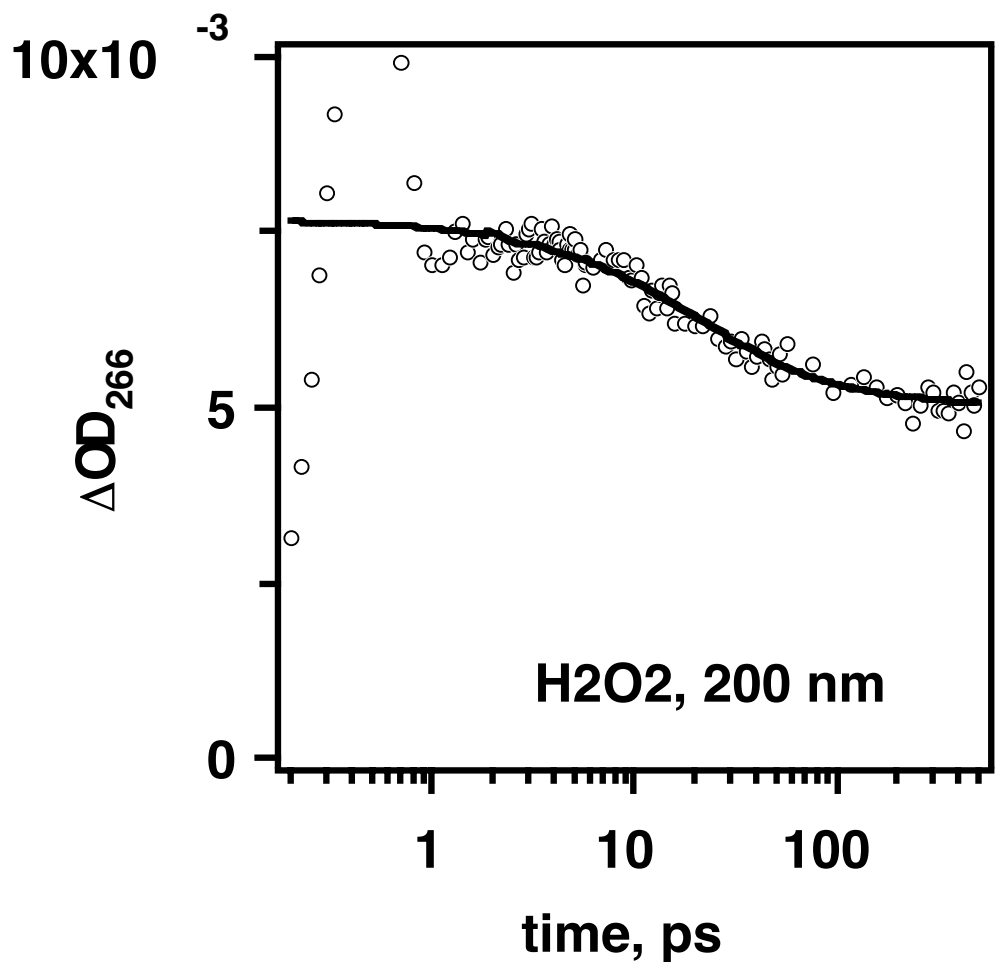

Figure 1.